\documentclass[journal=jacsat,manuscript=article]{achemso}

\usepackage{chemformula} 
\usepackage[T1]{fontenc} 
\usepackage{amsfonts}
\usepackage{amssymb}
\usepackage{color}

\author{Vasily G. Artemov}
\affiliation[Skoltech]
{Center for Energy Science and Technology, Skolkovo Institute of Science and Technology, 121205 Moscow, Russia}
\email{v.artemov@skoltech.ru}
\author{Alexander Ryzhov}
\affiliation[Skoltech]
{Center for Energy Science and Technology, Skolkovo Institute of Science and Technology, 121205 Moscow, Russia}
\author{Emma Carlsen}
\affiliation[BYU]
{Department of Chemistry and Biochemistry, Brigham Young University, 84602 Provo, Utah, USA}
\author{Pavel O. Kapralov}
\affiliation[RQC]
{Russian Quantum Center, 143025 Moscow, Russia}
\author{Henni Ouerdane}
\affiliation[Skoltech]
{Center for Energy Science and Technology, Skolkovo Institute of Science and Technology, 121205 Moscow, Russia}

\title {Non-rotational Mechanism of Polarization in Alcohols}

\begin{document}

\begin{abstract}
Chemical polarity governs various mechanical, chemical and thermodynamic properties of dielectrics. Polar liquids have been amply studied, yet the basic mechanisms underpinning their dielectric properties remain not fully understood, as standard models following Debye's phenomenological approach do not account for quantum effects and cannot aptly reproduce the full dc-up-to-THz spectral range. Here, using the illustrative case of monohydric alcohols, we show that deep tunneling and the consequent intermolecular separation of excess protons and ``proton-holes'' in the polar liquids govern their static and dynamic dielectric properties on the same footing. We performed systematic ultrabroadband (0-10 THz) spectroscopy experiments with monohydric alcohols of different (0.4-1.6 nm) molecular lengths, and show that the finite lifetime of molecular species, and the proton-hole correlation length are the two principle parameters responsible for the dielectric response of all the studied alcohols across the entire frequency range. Our results demonstrate that a quantum non-rotational intermolecular mechanism drives the polarization in alcohols while the rotational mechanism of molecular polarization plays a secondary role, manifesting itself in the sub-terahertz region only.
\end{abstract}

\section{Introduction}
A polar liquid can be seen as a system of permanent electric dipoles interacting with one another. The frequency-dependent response of such a liquid to the perturbation of an oscillating electric field manifests itself via the dielectric relaxation. The interpretation of dielectric spectra yields information on the dipole relaxation time and mechanisms, which in turn provides a detailed understanding of the physical and chemical properties of polar liquids, including their molecular structures, interactions, and dynamics. Water and alcohols are two typical examples of polar liquids, but many more may be found in nature or synthesized. In particular, some chelate compounds like hemoglobin and chlorophyll play a vital role in biological systems, thus illustrating the importance of chemical polarity.

The polar character of alcohols is due to the presence of the characteristic hydroxyl group, $-$OH, bound to a carbon atom that forms simple covalent bonds with other groups of atoms. Just like water, alcohols have a relatively high dielectric constant and are good solvents. Their typical size and relatively low complexity make them interesting systems to study with dielectric spectroscopy. That is why over the past decades, the interaction of electric fields with alcohols has been the object of experimental and theoretical research  \cite{Partington1952,Mashimo1991,Joo1996,Wang2004,Edenberg2007,Tomsic2007,Merle2011,Gainaru2010,Gastegger2017,Clarke2018,Carignani2018}, which showed many other fundamental similarities with water~\cite{Sillren2014, Hansen2016}. These findings notably triggered the studies of alcohols in the glassy state - an experimentally inaccessible thermodynamic region for water \cite{Wang2005,Gerstner2012}, which led to the prediction of some supercooled \cite{Lunkenheimer2017} and confined \cite{Artemov2020a} water properties, as well as deeper insights into dielectric phenomena in protonic liquids \cite{Bohmer2014}. However, the understanding on the microscopic level of the response of alcohol molecules to an alternating electric field remains far from complete \cite{Dutt1990,Gainaru2010,Weingartner2004}.

Debye's early phenomenological models of the dielectric polarization assumed molecules as rigid dipoles following the external electric field direction, with a characteristic time $\tau$ reflecting the average effect of molecular correlations \cite{Debye1912,Debye1929,Debye1935}. In this approach the parametric relationship between the real and imaginary parts of the dielectric function in Cole-Cole diagrams \cite{Cole1941,Cole1942} differs from experimental data in the high-frequency domain, implying that important short-time scale physics is missed altogether. For alcohols, the discrepancy between the measured dielectric constants and those determined by calculation with the Debye formula using a gas-phase molecular dipole moment value of $\mu\approx 1.7$ D becomes significant at low temperatures \cite{Miles1929}. In addition, the Debye model does not account for steric, and entropic effects that influence the molecules' geometry as shown with isomeric octanols~\cite{Dannhauser1968}. Accounting for intermolecular correlations and local-field effects \cite{Onsager1936,Kirkwood1939,Frohlich1949} by introducing for instance the Kirkwood g-factor, and using improved models for the relaxation time \cite{Glarum1960,Fang1965,Cole1965} yields good theory-experiment agreement for the dielectric constants of alcohols \cite{Bohmer2014,Gaudin2019}. However, other dynamical processes, such as self-diffusion, protonic current, and high-frequency (terahertz) modes as well as their temperature dependencies, still lack clarity \cite{Sillren2014,Jensen2018,Yamaguchi2018,Maribo2013,Angell2000}.

As an alternative, one may consider a chain-like structure made of several $-$OH groups as the transient chain model for monohydric alcohols \cite{Gainaru2010}. Then, if $\mu=1.7$ D, the total of 6 aligned moments in a chain is $\sim 10$ D, which is large enough to account for the dielectric constant. Nonetheless, several problems remain: i/ there is no proton transfer in this chain model, hence no dc conductivity; ii/ the dielectric relaxation time is too short to allow for the alignment of a sizable number of dipoles, required for the experimentally observed dielectric constant (see Appendix); iii/ the thermal energy $k_{\rm B}T= 0.026$ eV is an order of magnitude larger than the dipole-dipole interaction energy $E_{\rm dd}=\mu^{2}/(4\pi\epsilon_0r^{3}) \sim 10^{-3}$ eV, with $\epsilon_0$ being the vacuum permittivity, so no long-lived chain-like structures are possible in alcohols on the time scale of dielectric relaxation; iv/ the shape of the dielectric relaxation function appears not to depend on the molecular length~\cite{Mashimo1991}, hence there is no broadening of the relaxation band with increase of the molecular size as one would expect from the molecular-reorientation mechanism; and v/the chemical shift of OH-peaks in $^1$H-NMR in alcohols\cite{Lomas2015} assumes intensive proton exchange between molecules, which is completely missing in the existing models. In fact, knowledge of the dipole moment $\mu$ of a single molecule only is insufficient to explain the polarization-related phenomena in alcohols. In particular, quantum effects that have been extensively identified in water and biological systems for decades, even at room temperature and above\cite{Meng2015, Salna2016, Lowdin1963, Wang2020}, have never been considered in alcohols.

In the present work, we perform a systematic ultra-broadband dielectric spectroscopy study on a series of monohydric alcohols of increasing carbon chain length. In comparison with previous studies \cite{Mashimo1991,Wang2004,Gainaru2010,Clarke2018,Merle2011,Edenberg2007,Tomsic2007,Carignani2018,Gastegger2017,Joo1996,Partington1952,Sillren2014,Hansen2016,Wang2005,Gerstner2012,Lunkenheimer2017,Bohmer2014,Dutt1990}, we expand the frequency domain down to the static conductivity, and up to the terahertz range, and we consistently perform the measurements for all the alcohols with the same experimental setup. From this data we unravel the preeminent role of proton intermolecular dynamics in the dielectric response across the whole frequency range, and show that both the static conductivity and the high-frequency dielectric polarization of alcohols can be explained by considering the charge separation over intermolecular distances induced by proton tunneling. We show that the formation of short-living ``excess proton-proton hole'' pairs ($\oplus-\ominus$ dipoles), whose diffusion-controlled dynamics yields both the high dielectric constant and the relatively high dc conductivity, leaves to the rotational relaxation mechanism a secondary role in the polarization of alcohols.

\section{Methods}

We used commercially available high-purity ($>$ 99.9\%) primary monohydric alcohols, which we classify according to their molecular length, characterized by the number of carbon atoms, $n$, in their hydrocarbon chain: methanol, ethanol, 1-propanol, 1-butanol, and decanol ($n = 1,2,3,4,10$). We used Keysight E4980A and N9917A impedance analyzers, operating in the parallel plate and coaxial probe modes, respectively. The real, $\epsilon'$, and imaginary, $\epsilon''$, parts of the dielectric function were obtained following the standard procedure \cite{Kremer2003}. The sample temperature was controlled by Peltier elements within 0.3 K accuracy in the 283 - 363 K temperature range. The terahertz spectra were adapted from Refs.~\cite{Yomogida2010,Sarkar2017}. In this way, we accumulated spectral data from 1 kHz to 10 THz.

\section{Results}

\begin{figure}
  \includegraphics[width=0.5\textwidth]{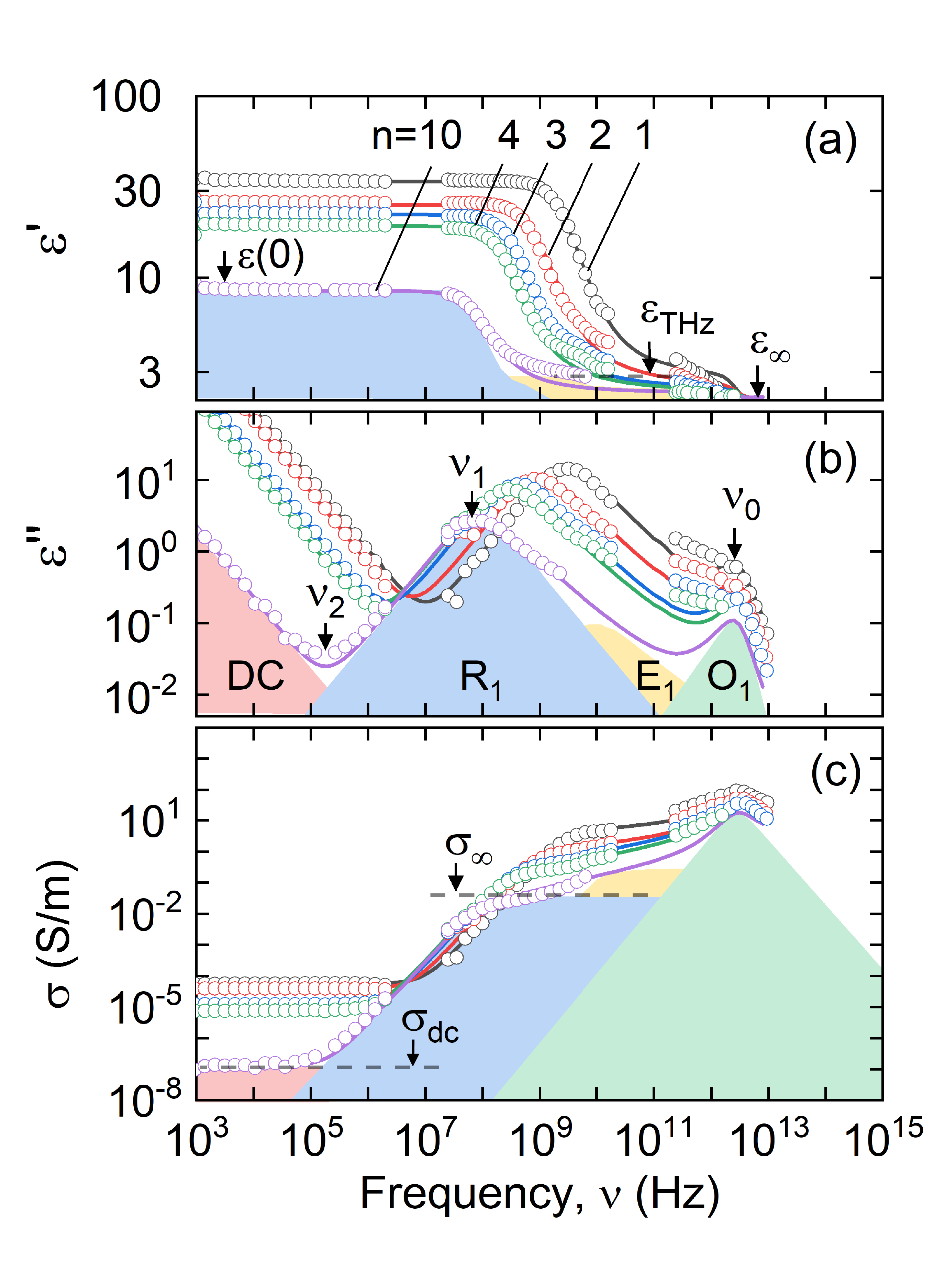}
  \caption{Broadband dielectric-terahertz spectra of monohydric alcohols for $n=1,2,3,4,10$. The panels (a) and (b) display the real $\epsilon'$ and imaginary $\epsilon''$ parts of the dielectric function; and (c) the dynamical conductivity $\sigma$. Experimental data are represented by open circles; continuous lines result from our model. Colored areas depict the main relaxation, R$_1$ (blue), its high-frequency satellite, E$_1$ (yellow), the static conductivity, DC (pink), and the far-infrared oscillation band, O$_1$ (green). The arrows indicate the characteristic parameters of the spectra, which are given in Table~\ref{tab1}, and their temperature dependencies are shown in Fig.~\ref{fig:3}.}
  \label{fig:1}
\end{figure}

Figure~\ref{fig:1} shows our measurement data of the alcohols' dielectric constants $\epsilon'(\nu)$, dielectric losses $\epsilon''(\nu)$, and dynamic conductivities $\sigma(\nu)=2\pi\epsilon''\epsilon_0\nu$. With molecular lengths ranging from 0.4 to 1.6 nm, the alcohol molecules stretch from almost spherical to rode-like geometries, retaining their dipole moment $\mu$. The static dielectric constant, $\epsilon'(0)$, direct-current conductivity, $\sigma_{\rm dc}$, high-frequency conductivity, $\sigma_{\infty}$, relaxation time, $\tau_{\rm r}$, and terahertz vibrational frequency, $\nu_0$, are also shown in Fig.~\ref{fig:1}, and their values given in Table~\ref{tab1}. The time $\tau_2$ in Fig.~\ref{fig:1}c corresponds to the ac-dc transition. The main relaxation band $R_1$ and its excess wing $E_1$ in Fig.~\ref{fig:1}b consistently shift towards lower frequencies as $n$ increases. The larger $\tau_{\rm r}$, the larger $\tau_2$, and the smaller $\epsilon(0)$ and $\sigma_{\rm dc}$. The frequency $\nu_0$ of the oscillator $O_1$ shown in Fig.~\ref{fig:1}c, varies fairly little with $n$.

\begin{table}
  \caption{Electrodynamic parameters of monohydric alcohols shown in Fig.~\ref{fig:1}, and derived parameters: $\tau_1$, $\tau_2$, $\Delta\epsilon_{\rm ionic}$, and $\Delta\epsilon_{\rm rot}$ (see text).}
  \label{tab1}
        \begin{tabular}{l|ccccc}
        \hline
        $n$ & 1 & 2 & 3 & 4 & 10\\
        \hline
        $\epsilon(0)$ & 34 & 26 & 22 & 19 & 8.5\\
        $\epsilon_{\rm THz}$ & 4.4 & 3.9 & 3.4 & 3.0 & 2.5\\
        $\epsilon_{\infty}$ & 2.2 & 2.2 & 2.2 & 2.2 & 2.2\\
        $\sigma_{\rm dc}$ ($\mu$S$\cdot$m$^{-1}$) & 57 & 37 & 12 & 7.4 & 0.13\\
        $\sigma_{\infty}$ (S$\cdot$m$^{-1}$) & 4.77 & 1.22 & 0.41 & 0.24 & 0.027\\
        $\nu_0$ (THz) & 2.89 & 2.57 & 2.39 & 2.20 & 2.05\\
        $\nu_1$ (GHz) & 2.95 & 0.90 & 0.43 & 0.29 & 0.08\\
        $\nu_2$ (MHz) & 9.95 & 5.31 & 1.99 & 1.30 & 0.17\\
        $\tau_1 = \tau_{\rm r} = (2\pi\nu_1)^{-1}$ (ps) & 54	& 177	& 370 & 549 & 1989\\
        $\tau_2 = \tau_{\rm m} = (2\pi\nu_2)^{-1}$ (ns) & 16 & 30 & 80 & 122 & 937\\
        $\Delta\epsilon_{\rm ionic} = \epsilon(0) - \epsilon_{\rm THz}$ (\%) & 87 & 86 & 85 & 84 & 74\\
         $\Delta\epsilon_{\rm rot} = \epsilon_{\rm THz} - \epsilon_{\infty}$ (\%) & 6.5 & 6.5 & 5.7 & 4.5 & 3.5\\
        \hline
        \end{tabular}
\end{table}

\begin{figure}
  \includegraphics[width=0.55\textwidth]{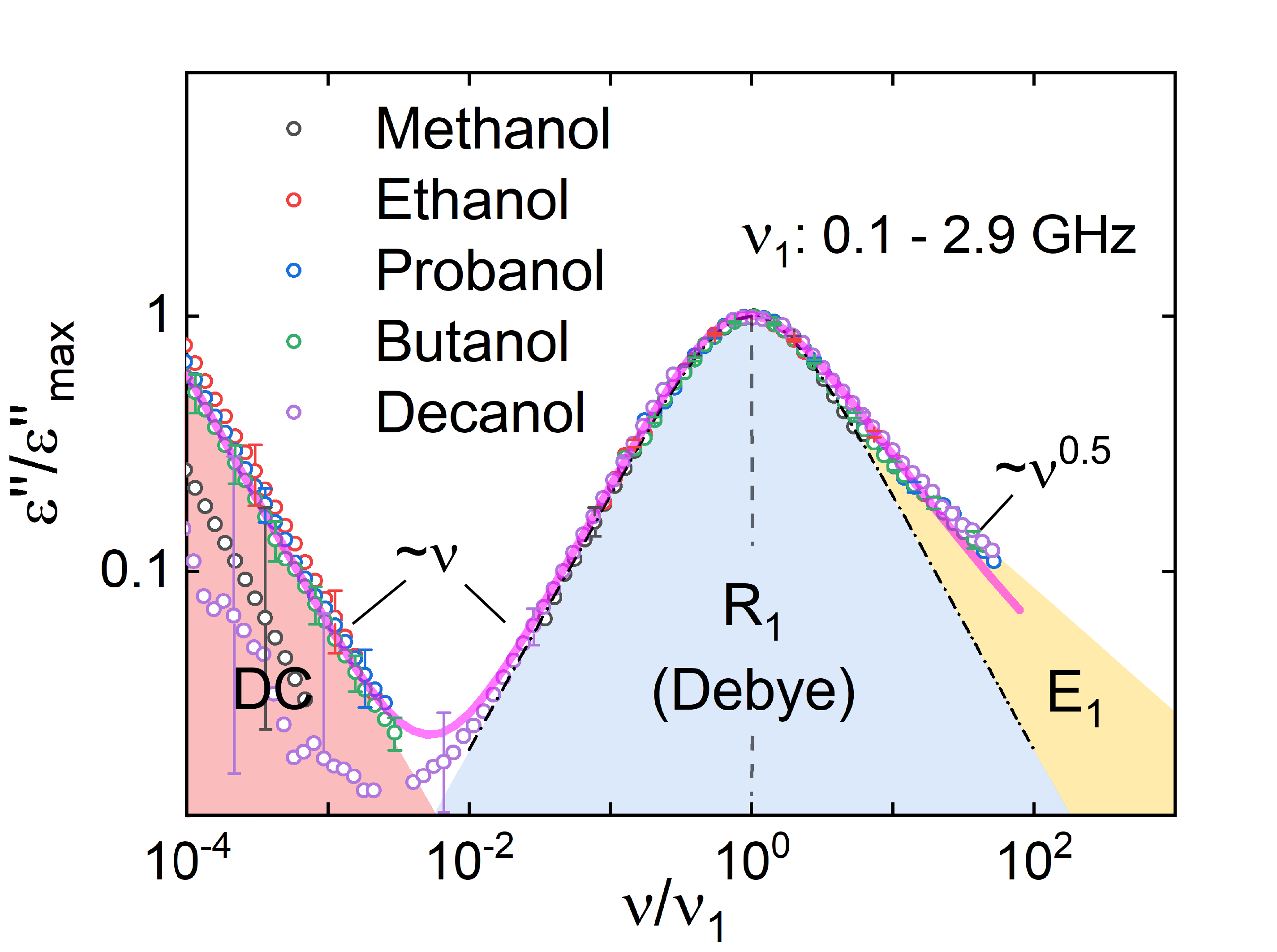}
  \caption{Normalized $\epsilon''/\epsilon_{\rm max}''$ spectra of monohydric alcohols. The dashed line corresponds to the Debye type of relaxation model based on molecular reorientation. Colored areas represent different regimes of the spectra: main relaxation (blue), its high frequency satellite (yellow), and the static conductivity (pink). The magents line is computed from our model and shows very good agreement with experimental data (circles) and some discrepancy with decanol only, but within the error bars.}
  \label{fig:2}
\end{figure}

Figure~\ref{fig:2} shows part of the dielectric losses data of Fig.~\ref{fig:1}b in the 1 kHz to 0.1 THz range. For each $n$ considered, the functions $\epsilon''(\nu)$ are normalized to their corresponding maximum value $\epsilon''_{\rm max}$ of the relaxation $R_1$ at corresponding frequency $\nu_1$ given in Table~\ref{tab1}. All the normalized $\epsilon''(\nu)/\epsilon''_{\rm max}$ spectra collapse into a unique master curve irrespective of their molecular aspect ratio (effective diameter/molecule length) for all alcohols considered in our work. This strongly suggests that the same microscopic mechanisms govern the dielectric losses for all alcohol studied. Note that the high-frequency wings $E_1$ and the dc conductivity (yellow and pink areas in Fig.~\ref{fig:2}), are also scaled. This implies that they are governed by a common molecular mechanism, which is the same for the main relaxation band $R_1$, again irrespective of the molecule geometries. Data accuracy for decanol is the object of a separate discussion (see Appendix).

\begin{figure}
  \includegraphics[width=0.6\textwidth]{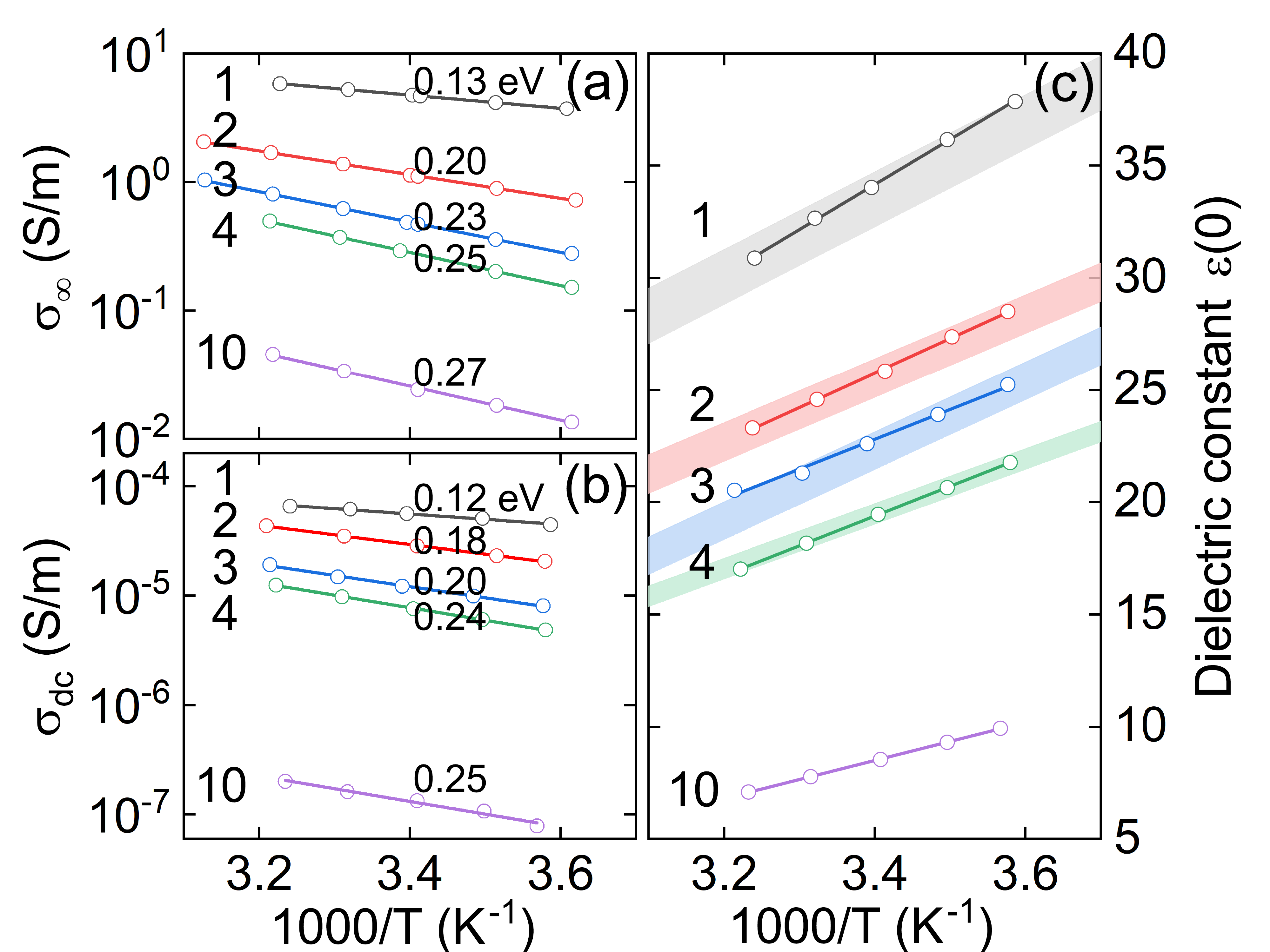}
  \caption{Temperature dependencies of (a) high-frequency conductivity, $\sigma_{\infty}$; (b) low-frequency static conductivity, $\sigma_{\rm dc}$; and (c) dielectric constant, $\epsilon(0)$. Color lines are fit according to Arrhenius law and Curie-Weiss law. Parameters are in Table 2. Numbers near curves are alcohols' ordinal numbers. Figures shown are activation energies in eV.}
  \label{fig:3}
\end{figure}

Figure~\ref{fig:3}, panels (a) and (b), show the temperature dependencies of $\sigma_{\rm dc}$ and $\sigma_{\infty}$, both in accord with the Arrhenius law: $\sigma_{\rm dc/\infty}=\sigma_{\rm dc/\infty}^{(0)}\exp(-E_{\rm dc/\infty}/k_{\rm B}T)$. We observe a systematic increase of the diffusion activation energy $E$ from 0.13 eV (methanol) to 0.27 eV (decanol) with increasing $n$. All alcohols studied show nearly identical trends of increasing $\sigma_{\rm dc}$ and $\sigma_{\infty}$. We also observe that for a given $n$, the activation energies are nearly the same when calculated from both the dc and high-frequency conductivity values (Table~\ref{tab2}). This property implies the same driving mechanism for both high- and low- frequency conductivities, which relates to self-diffusion of molecules. Note that the dc conductivities $\sigma_{\rm dc}$ measured in this study are lower than those obtained in \cite{Prego2000,Prego2003}. Further, our data clearly show that the activation energies $E$ increase with $n$, whereas the previous studies reported a constant value $E=0.16$ eV for all alcohols \cite{Prego2000,Prego2003}. This discrepancy may be caused by the different purity of samples: our samples have higher purity, and show consistent behavior of the conductivity, correlated with self-diffusion, which lends credibility to the reliability of our measurement data. Figure~\ref{fig:3}c shows the temperature dependencies of the static dielectric constant $\epsilon(0)$. Each curve follows a Curie-Weiss type of law: $\epsilon(0)\equiv 1+A_n/T$, with $A_n$ being an alcohol-specific variable explained further below. Note that $\epsilon(0)$ has the same nature as the main Debye relaxation $R_1$, because the latter gives a contribution $\Delta\epsilon_{\rm ionic}$ up to 90\% to $\epsilon(0)$ (see Table 1): the amplitude and position of the main relaxation $R_1$ determine the dielectric constant.

\begin{table}
\caption{Transport parameter of alcohols: $D$ is the diffusion coefficient \cite{Pratt1977,robb1982}, and $E$ its activation energy in eV; $\sigma_{\rm dc}^{(0)}$, $\sigma_{\infty}^{(0)}$, $E_{\rm dc}$, and $E_{\infty}$, are pre-exponential factors, and activation energies of direct current, and high frequency conductivities, respectively.}
\label{tab2}
        \begin{tabular}{l|ccccc}
        \hline
        $n$ & 1 & 2 & 3 & 4 & 10\\
        \hline
        $D$ (nm$^2$$\cdot$ns$^{-1}$) & 2.44 & 1.16 & 0.60 & 0.50 & 0.12\\
        $E$ (eV) & 0.13(3) & 0.20(1) & 0.24(1) & 0.25(1) & 0.26(1)\\
        $\sigma_{\rm dc}^{(0)}$ (S$\cdot$m$^{-1}$) & 6500 & 45100 & 32200 & 96300 & 2510\\
        $E_{\rm dc}$ (eV) & 0.12(2)	& 0.18(2)	& 0.20(3) & 0.24(2) & 0.25(3)\\
        $\sigma_{\infty}^{(0)}$ (S$\cdot$m$^{-1}$) & 4.77 & 1.22 & 0.41 & 0.24 & 0.027\\
        $E_{\infty}$ (eV) & 0.13(2) & 0.20(2) & 0.23(2) & 0.25(2) & 0.27\\
        
        \hline
        \end{tabular}
\end{table}

\begin{table}
\caption{\label{tab3}Structural parameters of alcohols: $d$ and $L$ are proton-holes pairs diffusion lengths in nm (see text); $N_{\rm m}$, $N_{\rm i}$, and $N_{\rm dc}$, are molar concentrations of alcohol molecules, short- and long-lived proton-hole pairs; $\alpha=N_i/(N_{\rm i}+N_{\rm m})$ is the degree of ionization; $pK_{\rm a}$ and $pK_{\rm a}$ (model) are dissociation constants \cite{Ugur2014}, and those calculated in our study, respectively.}
        \begin{tabular}{l|ccccc}
        \hline
        $n$ & 1 & 2 & 3 & 4 & 10\\
        \hline
        $d$ (nm) & 0.50 & 0.57 & 0.62 & 0.66 & 0.84\\
        $L$ (nm) & 40.6 & 36.6 & 42.8 & 47.4 & 108.0\\
        $\tau_{\rm OH}$ (ps) & 51 & 140 & 320 & 436 & 3207\\
        $N_{\rm m}$ (M)& 24.7	& 17.2	& 13.3 & 10.9 & 5.25\\
        $N_{\rm i}$ (M) & 0.5 & 0.3 & 0.2 & 0.1 & 0.05\\
        $N_{\rm dc}\times10^6$ (M) & 6.2 & 8.5 & 5.3 & 3.9 & 0.32\\
        $\alpha$ & 0.020 & 0.017 & 0.015 & 0.010 & 0.009\\
        $pK_{\rm a}$ (model) & 15.8 & 15.4 & 15.7 & 15.9 & 16.1\\
        $pK_{\rm a}$ & 15.5 & 15.5 & 16.1 & 16.1 & -\\
         \hline
        \end{tabular}
\end{table}

\section{Discussion}

As we experimentally observe uniform polarization dynamics of alcohols, we can unravel its underpinning physical mechanism. Accounting for quantum effects in liquids may be of importance to explain some of their properties \cite{Miller2005,Habershon2009,Novikov2013}. The strongly delocalized nature of the excited vibrational states of the O-H stretch vibration, as shown by ab initio path integral simulations, are responsible for the mechanism of proton transfer at room temperature\cite{Bakker2002}. Because water and alcohols share some common properties, we assume that proton intermolecular tunneling, a process well-identified in water \cite{Meng2015,Artemov2020}, also occurs in alcohols.

\begin{figure}
  \includegraphics[width=0.65\textwidth]{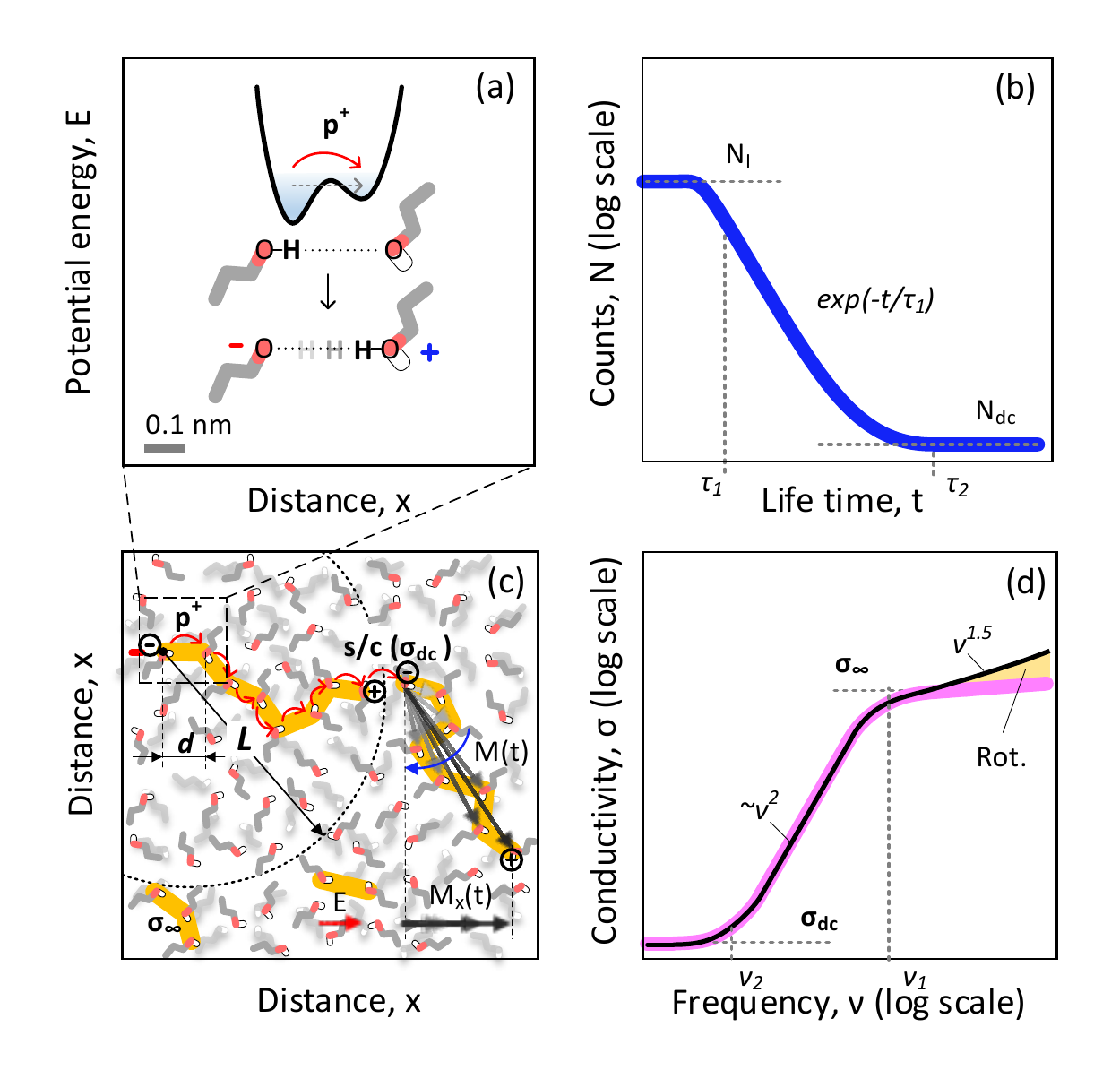}
  \caption{Electrodynamics of primary alcohols. (a): proton potential energy barrier between two neighboring molecules; (b) ionic species lifetime distribution with $\tau_1$ the ions characteristic lifetime, and $\tau_2$ the molecules lifetime; (c) microscopic structure explaining the ac- and dc- conductivity of alcohols; (d) dielectric response in terms of conductivity. The black line corresponds to the experimental data trend, the magenta line is our model, and the yellow area represents the part where experiment and model diverge, which has been associated with rotational component of polarization.}
  \label{fig:4}
\end{figure}

Figure~\ref{fig:4} is a depiction of our alcohol model. Fig.~\ref{fig:4}a represents the double-well potential formed by two bounded molecules. The gray tails represent H(CH$_2$)$_n$ alkyl groups (for alcohols) or just one hydrogen atom H (for water). Each proton of the OH group has a non-zero probability to tunnel through the potential barrier. Such hydrogen bridges, where a proton with electric charge $q$, is delocalized between the molecules, continuously form and break on the subpicosecond timescale \cite{Mazur2015}. Note that the relaxation time $\tau_{\rm r}$ (Table~\ref{tab1}) is larger. Proton tunneling requires two oxygen atoms to align exactly along the proton transfer line, which occurs on the 1-2 ps time scale \cite{Markovitch2008}. When the bond breaks, the proton can detach from the parent molecule, and forms a proton-hole pair, or two ionic species, RO$^{\ominus}$ and ROH$_2^{\oplus}$ \cite{Pietropaolo2008}. The energy difference between molecular and ionic states results from collective molecular effects in the adiabatic regime \cite{Lin2010}. The ions have been shown to be just 2 kcal/mol (0.09 eV) higher in energy than neutral molecules\cite{Geissler2001} (see potential minima in Fig.~\ref{fig:4}a). The majority of newly-formed ions, however, recombine within 1 ps \cite{Hassanali2013}, but some of them can live much longer. This lifetime is expected to be higher in alcohols because the probability of proton transfer is lower due to the increase of the distance between OH groups.

Figure~\ref{fig:4}b shows the time distribution of ionic species concentration: $N(t)=N_{\rm i}\exp(-t/\tau_1)+N_{\rm dc}$, where $N_{\rm i}$ is the instantaneous concentration of all ionic species (both short- and long-lived), $\tau_1$ is the characteristic time of excess proton state (proton-hole pair lifetime), and $N_{\rm dc}$ is the concentration of those ionic species, which reach the percolation threshold (Fig.~~\ref{fig:4}b) and show dc conductivity. Figure~\ref{fig:4}d shows the conductivity spectrum given by $\sigma(\nu) = (Dq^2/k_{\rm B}T)\int N(t)\exp(-i2\pi\nu t){\rm d}t$, and which, for $N_{\rm i}\gg N_{\rm dc}$ reduces to: 
\begin{equation}\label{sigmanu}
\sigma(\nu) = \frac{\displaystyle \sigma_{\rm dc}+(2\pi\nu)^2\tau_1^2\sigma_{\infty}}{\displaystyle 1+(2\pi\nu)^2\tau_1^2}
\end{equation}
\noindent where $\sigma_{\rm dc}=q^2N_{\rm dc}D/k_{\rm B}T$, $\sigma_{\infty}=q^2N_{\rm i}D/k_{\rm B}T$. Here excess protons and molecules have the same frequency-independent diffusion coefficient $D$  (Table~\ref{tab2}), as shown by tracer-diffusion experiments\cite{Hawlicka1995}. We associate this fact with the ``proton-sticking'' mechanism, when a proton, while transferring between the molecules, still stays attached to one or another molecule and thus has the same diffusion coefficient.

Equation~\eqref{sigmanu} reproduces well the experimental spectra of alcohols up to 0.1 THz (see the magenta curve in Figs.~\ref{fig:1}c and \ref{fig:4}d). For $\nu\rightarrow\infty$ it gives the $\sigma_{\infty}$ plateau, and for $\nu\rightarrow 0$ the $\sigma_{\rm dc}$ plateau. These two plateaus correspond to the same mechanism of proton-hole separation, with equal activation energies $E$, but the respective conductivity levels are different because if $t < \tau_1$ all pairs contribute; if $t > \tau_1$ only the long-lived pairs contribute. Comparing Eq.~\eqref{sigmanu} with the experimental spectrum in Fig.~\ref{fig:1}, we find that $\tau_{\rm r}$ (main Debye relaxation $R_1$) coincides with $\tau_1$ (proton-hole pairs lifetime). The ac-dc transition time equals $\tau_2=\tau_1\sqrt{\sigma_{\infty}/\sigma_{\rm dc}}$. The model \eqref{sigmanu} assumes that the polarization in alcohols is due to translational diffusion of unbound charges, namely excess protons and proton holes ($\oplus-\ominus$ dipoles) with lifetimes determined by separation and recombination. The proton-hole dynamical structure is similar to a plasma in the frame of reference of the globally neutral molecular network.

Figure~\ref{fig:4}c illustrates our proposed mechanism. Yellow lines show excess protons trajectories (``wires''). A spontaneous relative displacement of charges (protons and holes) changes the dipole moment $M(t)=q\Delta r(t)$ (black arrows) and its projection $M_x(t)=q\delta x(t)$ on the $x$-axis along the external electric field $E$. Note that the position of the proton hole is shown fixed for simplicity. The corresponding polarization $P_x(t)=\langle q\sum M_x(t)\rangle/V$ determines the dielectric function $\epsilon(\nu) = 1 +P_x(\nu)/\epsilon_0E$ and the dynamical conductivity $\sigma(\nu) = 2i\pi\nu\epsilon(\nu)\epsilon_0$. The lifetime distribution (Fig.~\ref{fig:4}b) imposes the wire length $l(t)=\sqrt{2Dt}$. The unit-length segment $l=d$ corresponds to $t=\tau_{\rm OH}$, and is equal to the distance $d=0.5\times (3/(4\pi N_m))^{1/3}$ between OH groups of randomly oriented molecules  (Table~\ref{tab3}). The time $\tau_{\rm OH}$ coincides with $\tau_1$. The second length $L$ represents the minimum length of proton wire required for the short-circuit (see 's/c' in Fig.~\ref{fig:4}c), or dc conduction $\sigma_{\rm dc}$. The corresponding time $t=\tau_2$, represents the moment when the mean-square-displacement spheres of protons start to overlap. The concentration $N_{\rm dc}$ of ionic species with lifetime $\tau_2$ equals $N_{\rm dc} = 3/(4\pi L^3)$, and corresponds to the dissociation constant $pK_{\rm a}$ (Table~\ref{tab3}). Note that the percolation threshold does not depend on the sample thickness \cite{Balberg1984}.

We can now derive the equation for the static dielectric constant, which according to the Debye formula reads $\epsilon(0) = \epsilon_{\rm THz} + \sigma_{\infty}\tau_{\rm r}/\epsilon_0$, from which we get: 
\begin{equation}\label{eps0}
\epsilon(0) = \epsilon_{\rm THz} + \frac{q^2N_{\rm m}^{1/2}}{k_{\rm B}T\epsilon_0} \left(\frac{9}{2\pi^2 {\rm K}_a^{1/2}}\right)^{1/3}
\end{equation}
\noindent where $\epsilon_{\rm THz}$ is shown in Fig.~\ref{fig:1}a, and $K{\rm_a}=N_{\rm dc}^2/N_{\rm m}$ is the dissociation constant (Table~\ref{tab2}). Note that the molecular concentration $N_{\rm m}$ represents here the concentration of OH groups. Equation~\eqref{eps0} fits well the experimental data including temperature dependencies (see lines in Fig.~\ref{fig:3}c). In contrast to the formula of the Debye-Onsager-Fr\"ohlich-Kirkwood approach \cite{Onsager1936,Kirkwood1939,Frohlich1949}, Eq.~\eqref{eps0} includes only one material-specific parameter, the dissociation constant $K_{\rm a}$, and does not require the knowledge the single-molecule dipole moment $\mu$. Therefore, within our model, the proton-hole dipole moments are responsible for polarization in alcohols.

Our model gives a relatively high instantaneous concentration $N_{\rm i}$ of excess protons ($\alpha \lesssim 2$\% - see Table~\ref{tab3}). This does not contradict the $pK_{\rm a}$ because only a small part of $N_{\rm i}$ actually contributes to the static conductivity, while $N_{\rm i}$ contributes fully to the high-frequency dielectric response. The concentration of short-lived ions of about 2\% of all alcohol molecules, assumes a potential barrier $\Delta E=-\ln(0.02)k_{\rm B}T \approx$ 0.1 eV (2.3 kcal/mol). This value is close to that obtained for water\cite{Geissler2001}.

Proton transport is often associated with a large kinetic isotope effect; however, it may not be clearly revealed in complicated multi-step reactions driving the dielectric relaxation, for which the measuring rates are much slower than the tunnelling timescale\cite{Salna2016}. Thus, the fact that both the relaxation time and the dielectric constant of methanol-h and methanol-d are very similar\cite{Davidson1957} does not preclude proton transfer. The relaxation time in our mode is the lifetime of RO$^{\ominus}-$H$^{\oplus}$(D$^{\oplus}$) ionic pairs, which is determined by the probability of proton tunneling and does not depend on the isotopic substitution. On the contrary, the rotational mechanism assumes a red shift of the relaxation band at isotopic substitution, which, was not observed\cite{Davidson1957}. This fact allows one to discriminate between rotation of dipoles and translation of free charges. Therefore, the translational mechanism is better suited for the experimental data explanation that the previously suggested rotational mechanism.

One can see in Figs.~\ref{fig:1}, and~\ref{fig:4}d that the suggested model and experimental data diverge in the terahertz frequency region. The contribution of this part to the static dielectric constant (see Table~\ref{tab1}) is $\Delta\epsilon_{\rm rot} = \epsilon_{\rm THz} - \epsilon_{\infty}$ and constitutes from 4 to 7 \% of the $\epsilon(0)$ value. We associate this excess absorption with molecular-rotation dynamism of polarization, in wich the molecular dipole moments follow the moving proton/hole with some delay. This fact explains the synchronism of the main relaxation band, R$_1$, and its excess wing, E$_1$ (as obvious from Fig.~\ref{fig:2}). However, the contribution of the latter is significantly smaller than commonly assumed for alcohols, and the corresponding part of the dielectric constant can be explained without the use of artificial overpolarization of molecular dipoles.

We associate the $O_1$ band (Fig.~\ref{fig:1}c) to the excess proton oscillation in the double-well potential. Indeed, the oscillator frequency $\nu = (2\pi x)^{-1}\sqrt{2\Delta E/m_{\rm p}} \approx$ 2.5 THz, where $m_{\rm p}$ is the proton mass, $x\approx$ 0.2 nm is average O$-$O distance, and $\Delta E\approx$ 0.2 eV is the energy barrier \cite{Sedov2012}, is close to $\nu_0$. The lifetime of the excess proton state (the half-width of the $\nu_0$ band) is equal to 0.5 ps, which is close to the known lifetime of the transition state of proton between molecules ~\cite{Thamer2015,Gainaru2011}. Moreover, this peak is temperature independent, as well as connected with infrared OH-stretch vibration\cite{Grechko2018}; thus, it is in line with the idea of proton tunneling. In light of our analysis, the mobile excess proton states of high concentration $N_{\rm i}$ but short lifetime, cause the local heterogeneity of alcohols, detected experimentally \cite{Guo2003,Gabriel2018}, and provoke the concomitant rearrangement of the chain structures observed in \cite{Singh2012}.

\section{Conclusion}

We experimentally measured the ultra-broadband dielectric response of five monohydric alcohols, and processed the data with a consistent methodology. We found similarities across the various alcohols, which are not fully explained by existing polarization models. Introducing excess protons and holes with exponential lifetime distributions, as a consequence of proton tunnelling, we propose a non-rotational polarization mechanism, which accounts on the same footing for static (dc conductivity and dielectric constant) and dynamic (dielectric relaxation and high frequency conductivity) effects as microscopically connected phenomena. Our model permits a consistent spectral data fitting across alcohols and provides a physical explanation to support observed data, giving new insights into the molecular dynamics of polar liquids. Our work shows that proton transport is fully consistent with dielectric spectroscopy data, but our model and interpretation still need to be tested using other experimental techniques and different types of protic liquids. We thus hope that this research will stimulate further experimental and theoretical activities on this front.

\begin{acknowledgement}
The Authors acknowledge support of the Skoltech NGP Program (Skoltech-MIT joint project). E.C. also acknowledges the Skoltech Global Campus Program.
\end{acknowledgement}

\section{Appendix}

\subsection{Samples preparation}
For our experiments we used five primary alcohols listed in Table~\ref{tab:alcohols}. All samples were purchased from the same manufacturer (Sigma-Aldrich), except ethanol (Fisher Chemical), and used as received. We used the samples with the highest available purity, i.e. 99.9\%, except for decanol with only 98\% purity, which explains the dispersion of the corresponding data on dc conductivity (see Fig. 2 of the main text). During the measurements, we found that if part of the probe is in contact with air, absorbed water vapor might affect the dielectric parameters of alcohols. However, this change only applies to static conductivity. The rest of the spectrum, above a few megahertz, was absolutely unaffected by atmospheric moisture over the time interval of the measurements, typically around one hour. Nevertheless, the influence of impurities on the static conductivity of alcohols should in principle be accounted for when considering previous published data on dc conductivity [46,47] (main text), as the samples used in these works had lower purity (see Table~\ref{tab:alcohols}).

\begin{table}
\renewcommand\thetable{S1} 
\caption{List of alcohols used in this study in comparison with those from previous research.}
\begin{tabular}{l | c  c  c }
  \hline
 Alcohol &~ This work ~&~ Ref.~[46] ~&~ Ref.~[47] \\
 \hline
 ~&~&~&~\\
 Methanol &~ Sigma-Aldrich, 99.9\% ~&~ Fluka, 99.8\% ~&~ Panreac, 99.8\% \\
 ~&~&~&~\\
 Ethanol &~ Fisher Chemical, 99.9\% ~&~ Fluka, 99.8\% ~&~ Fluka, 99.8\% \\
 ~&~&~&~\\
 1-Propanol &~ Sigma-Aldrich, 99.9\% ~&~ Fluka, 99.5\% ~&~ Fluka, 99.5\% \\
 ~&~&~&~\\
 1-Butanol &~ Sigma-Aldrich, 99.9\% ~&~ Fluka, 99.5\% ~&~ Fluka, 99.5\% \\
 ~&~&~&~\\
 1-Decanol &~ Sigma-Aldrich, 98\% ~&~ - ~&~ - \\
 \hline
 \end{tabular}
\label{tab:alcohols}
\end{table}
	
\subsection{Measurement procedure}
The low- (megahertz), and high-frequency (gigahertz) measurements were performed using different sample holders and measuring cells. Alcohols were placed into measurement cells using Finnpipette renewed after each usage. The cell was washed with ethanol before each measurement and stored in a liquid to be measured for 5 minutes, heated up to 40$^{\circ}$C. The dc measurements cell was thermally stabilized using acetone as a thermal conductive liquid, Peltier cooler for temperature control, and Pt1000 thermocouples for accurate temperature measurements. For the gigahertz region measurements, we used a copper bath, in which the sample in the glass beaker was tightly inserted. The open-end coaxial probe was applied at the liquid-air interface. At low frequencies, we used a cylindrical Teflon cell with two round-flat gold electrodes of 1 cm$^{2}$ each and separation about 1 mm. The complex impedance was measured by the four-electrode method [41] (main text), as each alcohol was passing through the space between the electrodes at a constant flow rate controlled by the peristaltic pump. 

The measuring rms voltage $V_{\rm ac}$=100 mV and signal intensity 45 dB were chosen well below the electrolysis stability threshold, 1.23 V, and such that the sample temperature and chemical composition remain stable. The low- and high-frequency measurements show comparable data for the static dielectric constant, and also for the static conductivity, which confirms the validity of our experimental approach. The real, $\epsilon'$, and imaginary, $\epsilon''$, parts of the complex dielectric permittivity are calculated from the measured complex impedance $Z^*=Z'+Z''$ with:

\begin{eqnarray}
  \nonumber
  \epsilon'(\omega) & = & \frac{1}{C_0}~\frac{-Z''}{\left({Z'}^2 + {Z''}^2\right)\omega}\\
  \nonumber
  \epsilon''(\omega) & = & \frac{1}{C_0}~\frac{Z'}{\left({Z'}^2 + {Z''}^2\right)\omega}
\end{eqnarray}

\noindent where $C_0$ is the capacitance of the empty cell. The dynamical conductivity is then obtained with its definition: $\sigma(\omega)=\epsilon''(\omega)\epsilon_0\omega$.

\subsection{Self-diffusion and dielectric relaxation}
The self-diffusion coefficient of molecules in liquids can be measured independently by two main methods, which give similar results: isotopic substitution, and spin-echo NMR. The data on self diffusion of oxygen, and hydrogen atoms in alcohols, obtained by these methods are available in Refs.~[46,47] (main text). For our study, it is important to know the mean square displacement $x$ of atoms on the timescale of the dielectric relaxation time $\tau_{\rm r}$, which can be obtained using the self-diffusion coefficients $D$ calculated with the Smoluchowski formula $x=(6D\tau_{\rm r})^{1/2}$. Values are given in Table S2 for all alcohols considered in our work. As the molecular diffusion in liquids demonstrates hoping-like behavior~\cite{Frenkel1946}, it would be informative to compare $x$ with the distance $d$ between centers of molecules. As one can see from Table~\ref{tab:coefs}, $x>d$; hence, over the relaxation time $\tau_{\rm r}$, each molecule covers at least one intermolecular distance as it moves, or, in other words, changes its local environment. This means that inasmuch as dielectric relaxation in associated liquids (including alcohols) is a collective phenomenon, no cluster-like structures made of several molecules can explain the dielectric relaxation, as they simply cannot last sufficiently long for such a relatively long time as $\tau_{\rm r}$.

\begin{table}[h!]
\renewcommand\thetable{S2} 
\caption{Self-diffusion and dielectric relaxation parameters of primary alcohols.}
\begin{tabular}{l|ccccc}
\hline
$n$ & 1 & 2 & 3 & 4 & 10\\
\hline
$\tau_{\rm r}$ (ps) & 54	& 177	& 370 & 549 & 1989\\
$D$ (nm$^2$$\cdot$ns$^{-1}$) & 2.44 & 1.16 & 0.60 & 0.50 & 0.12\\
$x$ (nm) & 0.88 & 1.11 & 1.16 & 1.28 & 1.21\\
$d$ (nm) & 0.50 & 0.57 & 0.62 & 0.66 & 0.84\\
\hline
\end{tabular}
\label{tab:coefs}
\end{table}

\bibliography{References}

\providecommand{\latin}[1]{#1}
\makeatletter
\providecommand{\doi}
  {\begingroup\let\do\@makeother\dospecials
  \catcode`\{=1 \catcode`\}=2 \doi@aux}
\providecommand{\doi@aux}[1]{\endgroup\texttt{#1}}
\makeatother
\providecommand*\mcitethebibliography{\thebibliography}
\csname @ifundefined\endcsname{endmcitethebibliography}
  {\let\endmcitethebibliography\endthebibliography}{}
\begin{mcitethebibliography}{74}
\providecommand*\natexlab[1]{#1}
\providecommand*\mciteSetBstSublistMode[1]{}
\providecommand*\mciteSetBstMaxWidthForm[2]{}
\providecommand*\mciteBstWouldAddEndPuncttrue
  {\def\EndOfBibitem{\unskip.}}
\providecommand*\mciteBstWouldAddEndPunctfalse
  {\let\EndOfBibitem\relax}
\providecommand*\mciteSetBstMidEndSepPunct[3]{}
\providecommand*\mciteSetBstSublistLabelBeginEnd[3]{}
\providecommand*\EndOfBibitem{}
\mciteSetBstSublistMode{f}
\mciteSetBstMaxWidthForm{subitem}{(\alph{mcitesubitemcount})}
\mciteSetBstSublistLabelBeginEnd
  {\mcitemaxwidthsubitemform\space}
  {\relax}
  {\relax}

\bibitem[Partington \latin{et~al.}(1952)Partington, Hudson, and
  Bagnall]{Partington1952}
Partington,~J.~R.; Hudson,~R.~F.; Bagnall,~K.~W. Self-diffusion of aliphatic
  alcohols. \emph{Nature} \textbf{1952}, \emph{169}, 583--584\relax
\mciteBstWouldAddEndPuncttrue
\mciteSetBstMidEndSepPunct{\mcitedefaultmidpunct}
{\mcitedefaultendpunct}{\mcitedefaultseppunct}\relax
\EndOfBibitem
\bibitem[Mashimo and Umehara(1991)Mashimo, and Umehara]{Mashimo1991}
Mashimo,~S.; Umehara,~T. Structures of water and primary alcohol studied by
  microwave dielectric analyses. \emph{J. Chem. Phys.} \textbf{1991},
  \emph{95}, 6257--6260\relax
\mciteBstWouldAddEndPuncttrue
\mciteSetBstMidEndSepPunct{\mcitedefaultmidpunct}
{\mcitedefaultendpunct}{\mcitedefaultseppunct}\relax
\EndOfBibitem
\bibitem[Joo \latin{et~al.}(1996)Joo, Jia, Yu, Lang, and Fleming]{Joo1996}
Joo,~T.; Jia,~Y.; Yu,~J.-Y.; Lang,~M.~J.; Fleming,~G.~R. Third-order nonlinear
  time domain probes of solvation dynamics. \emph{J. Chem. Phys.}
  \textbf{1996}, \emph{104}, 6089--6108\relax
\mciteBstWouldAddEndPuncttrue
\mciteSetBstMidEndSepPunct{\mcitedefaultmidpunct}
{\mcitedefaultendpunct}{\mcitedefaultseppunct}\relax
\EndOfBibitem
\bibitem[Wang and Richert(2004)Wang, and Richert]{Wang2004}
Wang,~L.~M.; Richert,~R. Dynamics of glass-forming liquids. IX. Structural
  versus dielectric relaxation in monohydroxy alcohols. \emph{J. Chem. Phys.}
  \textbf{2004}, \emph{121}, 11170--11176\relax
\mciteBstWouldAddEndPuncttrue
\mciteSetBstMidEndSepPunct{\mcitedefaultmidpunct}
{\mcitedefaultendpunct}{\mcitedefaultseppunct}\relax
\EndOfBibitem
\bibitem[Edenberg(2007)]{Edenberg2007}
Edenberg,~H.~J. The genetics of alcohol metabolism: Role of alcohol
  dehydrogenase and aldehyde dehydrogenase variants. \emph{Alcohol Research \&
  Health} \textbf{2007}, \emph{30}, 5--13\relax
\mciteBstWouldAddEndPuncttrue
\mciteSetBstMidEndSepPunct{\mcitedefaultmidpunct}
{\mcitedefaultendpunct}{\mcitedefaultseppunct}\relax
\EndOfBibitem
\bibitem[Tom\v{s}i\v{c} \latin{et~al.}(2007)Tom\v{s}i\v{c}, Jamnik,
  Fritz-Popovski, Glatter, and Vl\v{c}ek]{Tomsic2007}
Tom\v{s}i\v{c},~M.; Jamnik,~A.; Fritz-Popovski,~G.; Glatter,~O.; Vl\v{c}ek,~L.
  Structural properties of pure simple alcohols from ethanol, propanol,
  butanol, pentanol, to hexanol: Comparing Monte Carlo simulations with
  experimental SAXS data. \emph{J. Phys. Chem. B} \textbf{2007}, \emph{111},
  1738--1751\relax
\mciteBstWouldAddEndPuncttrue
\mciteSetBstMidEndSepPunct{\mcitedefaultmidpunct}
{\mcitedefaultendpunct}{\mcitedefaultseppunct}\relax
\EndOfBibitem
\bibitem[Merle \latin{et~al.}(2011)Merle, Wessling, and Nijmeijer]{Merle2011}
Merle,~G.; Wessling,~M.; Nijmeijer,~K. Anion exchange membranes for alkaline
  fuel cells: A review. \emph{J. Membr. Sci.} \textbf{2011}, \emph{377},
  1--35\relax
\mciteBstWouldAddEndPuncttrue
\mciteSetBstMidEndSepPunct{\mcitedefaultmidpunct}
{\mcitedefaultendpunct}{\mcitedefaultseppunct}\relax
\EndOfBibitem
\bibitem[Gainaru \latin{et~al.}(2010)Gainaru, Meier, Schildmann, Lederle,
  Hiller, R\"ossler, and B\"ohmer]{Gainaru2010}
Gainaru,~C.; Meier,~R.; Schildmann,~S.; Lederle,~C.; Hiller,~W.; R\"ossler,~E.;
  B\"ohmer,~R. Nuclear-magnetic-resonance measurements reveal the origin of the
  Debye process in monohydroxy alcohols. \emph{Phys. Rev. Lett.} \textbf{2010},
  \emph{105}, 258303--4\relax
\mciteBstWouldAddEndPuncttrue
\mciteSetBstMidEndSepPunct{\mcitedefaultmidpunct}
{\mcitedefaultendpunct}{\mcitedefaultseppunct}\relax
\EndOfBibitem
\bibitem[Gastegger \latin{et~al.}(2017)Gastegger, Behler, and
  Marquetand]{Gastegger2017}
Gastegger,~M.; Behler,~J.; Marquetand,~P. Machine learning molecular dynamics
  for the simulation of infrared spectra. \emph{Chem. Sci.} \textbf{2017},
  \emph{8}, 6924--6935\relax
\mciteBstWouldAddEndPuncttrue
\mciteSetBstMidEndSepPunct{\mcitedefaultmidpunct}
{\mcitedefaultendpunct}{\mcitedefaultseppunct}\relax
\EndOfBibitem
\bibitem[Clarke \latin{et~al.}(2018)Clarke, Tu, Levers, Br\"ohl, and
  Hallett]{Clarke2018}
Clarke,~C.~J.; Tu,~W.-C.; Levers,~O.; Br\"ohl,~A.; Hallett,~J.~P. Green and
  sustainable solvents in chemical processes. \emph{Chem. Rev.} \textbf{2018},
  \emph{118}, 747--800\relax
\mciteBstWouldAddEndPuncttrue
\mciteSetBstMidEndSepPunct{\mcitedefaultmidpunct}
{\mcitedefaultendpunct}{\mcitedefaultseppunct}\relax
\EndOfBibitem
\bibitem[Carignani \latin{et~al.}(2018)Carignani, C.~Forte, Ga\l{}azka,
  Massalska-Arod\'z, Geppi, and Calucci]{Carignani2018}
Carignani,~E.; C.~Forte,~E. J.-G.; Ga\l{}azka,~M.; Massalska-Arod\'z,~M.;
  Geppi,~M.; Calucci,~L. Dynamics of two glass forming monohydroxy alcohols by
  field cycling 1H NMR relaxometry. \emph{J. Mol. Liq.} \textbf{2018},
  \emph{269}, 847--852\relax
\mciteBstWouldAddEndPuncttrue
\mciteSetBstMidEndSepPunct{\mcitedefaultmidpunct}
{\mcitedefaultendpunct}{\mcitedefaultseppunct}\relax
\EndOfBibitem
\bibitem[Sillr\'en \latin{et~al.}(2014)Sillr\'en, Matic, Karlsson, Koza,
  Maccarini, Fouquet, G\"otz, Bauer, Gulich, Lunkenheimer, Loidl, Mattsson,
  Gainaru, Vynokur, Schildmann, Bauer, and B\"ohmer]{Sillren2014}
Sillr\'en,~P. \latin{et~al.}  Liquid 1-propanol studied by neutron scattering,
  near-infrared, and dielectric spectroscopy. \emph{J. Chem. Phys.}
  \textbf{2014}, \emph{140}, 124501--10\relax
\mciteBstWouldAddEndPuncttrue
\mciteSetBstMidEndSepPunct{\mcitedefaultmidpunct}
{\mcitedefaultendpunct}{\mcitedefaultseppunct}\relax
\EndOfBibitem
\bibitem[Hansen \latin{et~al.}(2016)Hansen, Kisliuk, Sokolov, and
  Gainaru]{Hansen2016}
Hansen,~J.~S.; Kisliuk,~A.; Sokolov,~A.~P.; Gainaru,~C. Identification of
  structural relaxation in the dielectric response of water. \emph{Phys. Rev.
  Lett.} \textbf{2016}, \emph{116}, 237601--5\relax
\mciteBstWouldAddEndPuncttrue
\mciteSetBstMidEndSepPunct{\mcitedefaultmidpunct}
{\mcitedefaultendpunct}{\mcitedefaultseppunct}\relax
\EndOfBibitem
\bibitem[Wang and Richert(2005)Wang, and Richert]{Wang2005}
Wang,~L.; Richert,~R. Debye type dielectric relaxation and the glass transition
  of alcohols. \emph{J. Phys. Chem. B} \textbf{2005}, \emph{109},
  11091--11094\relax
\mciteBstWouldAddEndPuncttrue
\mciteSetBstMidEndSepPunct{\mcitedefaultmidpunct}
{\mcitedefaultendpunct}{\mcitedefaultseppunct}\relax
\EndOfBibitem
\bibitem[Gerstner(2012)]{Gerstner2012}
Gerstner,~E. Liquids in no man's land. \emph{Nature Phys.} \textbf{2012},
  \emph{8}, 252--1\relax
\mciteBstWouldAddEndPuncttrue
\mciteSetBstMidEndSepPunct{\mcitedefaultmidpunct}
{\mcitedefaultendpunct}{\mcitedefaultseppunct}\relax
\EndOfBibitem
\bibitem[Lunkenheimer \latin{et~al.}(2017)Lunkenheimer, Emmert, Gulich,
  K\"ohler, Wolf, Schwab, and Loidl]{Lunkenheimer2017}
Lunkenheimer,~P.; Emmert,~S.; Gulich,~R.; K\"ohler,~M.; Wolf,~M.; Schwab,~M.;
  Loidl,~A. Electromagnetic-radiation absorption by water. \emph{Phys. Rev. E}
  \textbf{2017}, \emph{96}, 062607--10\relax
\mciteBstWouldAddEndPuncttrue
\mciteSetBstMidEndSepPunct{\mcitedefaultmidpunct}
{\mcitedefaultendpunct}{\mcitedefaultseppunct}\relax
\EndOfBibitem
\bibitem[Artemov \latin{et~al.}(2020)Artemov, Uykur, Kapralov, Kiselev,
  Stevenson, Ouerdane, and Dressel]{Artemov2020a}
Artemov,~V.~G.; Uykur,~E.; Kapralov,~P.; Kiselev,~A.; Stevenson,~K.;
  Ouerdane,~H.; Dressel,~M. Anomalously high proton conduction of interfacial
  water. \emph{J. Phys. Chem. Lett.} \textbf{2020}, \emph{11}, 3623--3628\relax
\mciteBstWouldAddEndPuncttrue
\mciteSetBstMidEndSepPunct{\mcitedefaultmidpunct}
{\mcitedefaultendpunct}{\mcitedefaultseppunct}\relax
\EndOfBibitem
\bibitem[B\"ohmer \latin{et~al.}(2014)B\"ohmer, Gainarua, and
  Richert]{Bohmer2014}
B\"ohmer,~R.; Gainarua,~C.; Richert,~R. Structure and dynamics of monohydroxy
  alcohols—Milestones towards their microscopic understanding, 100 years
  after Debye. \emph{Phys. Rep.} \textbf{2014}, \emph{545}, 125--195\relax
\mciteBstWouldAddEndPuncttrue
\mciteSetBstMidEndSepPunct{\mcitedefaultmidpunct}
{\mcitedefaultendpunct}{\mcitedefaultseppunct}\relax
\EndOfBibitem
\bibitem[Dutt \latin{et~al.}(1990)Dutt, Doraiswamy, Periasamy, and
  Venkataraman]{Dutt1990}
Dutt,~G.~B.; Doraiswamy,~S.; Periasamy,~N.; Venkataraman,~B. Rotational
  reorientation dynamics of polar dye molecular probes by picosecond laser
  spectroscopic technique. \emph{J. Chem. Phys.} \textbf{1990}, \emph{93},
  8498--8513\relax
\mciteBstWouldAddEndPuncttrue
\mciteSetBstMidEndSepPunct{\mcitedefaultmidpunct}
{\mcitedefaultendpunct}{\mcitedefaultseppunct}\relax
\EndOfBibitem
\bibitem[Weing\"artner \latin{et~al.}(2004)Weing\"artner, Nadolny, Oleinikova,
  and Ludwig]{Weingartner2004}
Weing\"artner,~H.; Nadolny,~H.; Oleinikova,~A.; Ludwig,~R. Collective
  contributions to the dielectric relaxation of hydrogen-bonded liquids.
  \emph{J. Chem. Phys.} \textbf{2004}, \emph{120}, 11692--11697\relax
\mciteBstWouldAddEndPuncttrue
\mciteSetBstMidEndSepPunct{\mcitedefaultmidpunct}
{\mcitedefaultendpunct}{\mcitedefaultseppunct}\relax
\EndOfBibitem
\bibitem[Debye(1912)]{Debye1912}
Debye,~P. Theorie der Dipolmomente der Molekeln. \emph{Physik. Z.}
  \textbf{1912}, \emph{13}, 97--100\relax
\mciteBstWouldAddEndPuncttrue
\mciteSetBstMidEndSepPunct{\mcitedefaultmidpunct}
{\mcitedefaultendpunct}{\mcitedefaultseppunct}\relax
\EndOfBibitem
\bibitem[Debye(1929)]{Debye1929}
Debye,~P. \emph{Polar Molecules}, 1st ed.; Chemical Catalog Co., Inc.: New
  York, 1929\relax
\mciteBstWouldAddEndPuncttrue
\mciteSetBstMidEndSepPunct{\mcitedefaultmidpunct}
{\mcitedefaultendpunct}{\mcitedefaultseppunct}\relax
\EndOfBibitem
\bibitem[Debye(1935)]{Debye1935}
Debye,~P. Dielektrische s\"{a}ttigung und behinderung der freien rotation in
  fl\"{u}ssigkeiten. \emph{Phys. Z.} \textbf{1935}, \emph{36}, 193--197\relax
\mciteBstWouldAddEndPuncttrue
\mciteSetBstMidEndSepPunct{\mcitedefaultmidpunct}
{\mcitedefaultendpunct}{\mcitedefaultseppunct}\relax
\EndOfBibitem
\bibitem[Cole and Cole(1941)Cole, and Cole]{Cole1941}
Cole,~K.~S.; Cole,~R.~H. Dispersion and absorption in dielectrics I.
  Alternating current characteristics. \emph{J. Chem. Phys.} \textbf{1941},
  \emph{9}, 341--351\relax
\mciteBstWouldAddEndPuncttrue
\mciteSetBstMidEndSepPunct{\mcitedefaultmidpunct}
{\mcitedefaultendpunct}{\mcitedefaultseppunct}\relax
\EndOfBibitem
\bibitem[Cole and Cole(1942)Cole, and Cole]{Cole1942}
Cole,~K.~S.; Cole,~R.~H. Dispersion and absorption in dielectrics II. Direct
  current characteristics. \emph{J. Chem. Phys.} \textbf{1942}, \emph{9},
  98--105\relax
\mciteBstWouldAddEndPuncttrue
\mciteSetBstMidEndSepPunct{\mcitedefaultmidpunct}
{\mcitedefaultendpunct}{\mcitedefaultseppunct}\relax
\EndOfBibitem
\bibitem[Miles(1929)]{Miles1929}
Miles,~J.~B. The dielectric constant and electric moment of some alcohol
  vapors. \emph{Phys. Rev.} \textbf{1929}, \emph{34}, 964--971\relax
\mciteBstWouldAddEndPuncttrue
\mciteSetBstMidEndSepPunct{\mcitedefaultmidpunct}
{\mcitedefaultendpunct}{\mcitedefaultseppunct}\relax
\EndOfBibitem
\bibitem[Dannhauser(1968)]{Dannhauser1968}
Dannhauser,~W.~J. Dielectric study of intermolecular association in isomeric
  octyl alcohols. \emph{Chem. Phys.} \textbf{1968}, \emph{48}, 1911--1917\relax
\mciteBstWouldAddEndPuncttrue
\mciteSetBstMidEndSepPunct{\mcitedefaultmidpunct}
{\mcitedefaultendpunct}{\mcitedefaultseppunct}\relax
\EndOfBibitem
\bibitem[Onsager(1936)]{Onsager1936}
Onsager,~L. Electric moments of molecules in liquids. \emph{J. Am. Chem. Soc.}
  \textbf{1936}, \emph{58}, 1486--1493\relax
\mciteBstWouldAddEndPuncttrue
\mciteSetBstMidEndSepPunct{\mcitedefaultmidpunct}
{\mcitedefaultendpunct}{\mcitedefaultseppunct}\relax
\EndOfBibitem
\bibitem[Kirkwood(1939)]{Kirkwood1939}
Kirkwood,~J.~G. The Dielectric Polarization of Polar Liquids. \emph{J. Chem.
  Phys.} \textbf{1939}, \emph{7}, 911--919\relax
\mciteBstWouldAddEndPuncttrue
\mciteSetBstMidEndSepPunct{\mcitedefaultmidpunct}
{\mcitedefaultendpunct}{\mcitedefaultseppunct}\relax
\EndOfBibitem
\bibitem[Fr\"ohlich(1949)]{Frohlich1949}
Fr\"ohlich,~H. \emph{Theory of Dielectrics}, 1st ed.; Oxford University Press:
  London, 1949\relax
\mciteBstWouldAddEndPuncttrue
\mciteSetBstMidEndSepPunct{\mcitedefaultmidpunct}
{\mcitedefaultendpunct}{\mcitedefaultseppunct}\relax
\EndOfBibitem
\bibitem[Glarum(1960)]{Glarum1960}
Glarum,~S.~H. Dielectric theory of polar liquids. \emph{J. Chem. Phys.}
  \textbf{1960}, \emph{33}, 1371--1375\relax
\mciteBstWouldAddEndPuncttrue
\mciteSetBstMidEndSepPunct{\mcitedefaultmidpunct}
{\mcitedefaultendpunct}{\mcitedefaultseppunct}\relax
\EndOfBibitem
\bibitem[Fang(1965)]{Fang1965}
Fang,~P.~H. Cole-Cole diagram and the distribution of relaxation times.
  \emph{J. Chem. Phys.} \textbf{1965}, \emph{42}, 3411--3413\relax
\mciteBstWouldAddEndPuncttrue
\mciteSetBstMidEndSepPunct{\mcitedefaultmidpunct}
{\mcitedefaultendpunct}{\mcitedefaultseppunct}\relax
\EndOfBibitem
\bibitem[Cole(1965)]{Cole1965}
Cole,~R.~H. Correlation theory of dielectric relaxation. \emph{J. Chem. Phys.}
  \textbf{1965}, \emph{42}, 637--643\relax
\mciteBstWouldAddEndPuncttrue
\mciteSetBstMidEndSepPunct{\mcitedefaultmidpunct}
{\mcitedefaultendpunct}{\mcitedefaultseppunct}\relax
\EndOfBibitem
\bibitem[Gaudin and Ma(2019)Gaudin, and Ma]{Gaudin2019}
Gaudin,~T.; Ma,~H. A molecular contact theory for simulating polarization:
  application to dielectric constant prediction. \emph{Phys. Chem. Chem. Phys}
  \textbf{2019}, \emph{21}, 14846--14857\relax
\mciteBstWouldAddEndPuncttrue
\mciteSetBstMidEndSepPunct{\mcitedefaultmidpunct}
{\mcitedefaultendpunct}{\mcitedefaultseppunct}\relax
\EndOfBibitem
\bibitem[Jensen \latin{et~al.}(2018)Jensen, Gainaru, Alba-Simionesco, Hecksher,
  and Niss]{Jensen2018}
Jensen,~M.~H.; Gainaru,~C.; Alba-Simionesco,~C.; Hecksher,~T.; Niss,~K. Slow
  rheological mode in glycerol and glycerol–water mixtures. \emph{Phys. Chem.
  Chem. Phys} \textbf{2018}, \emph{20}, 1716--1723\relax
\mciteBstWouldAddEndPuncttrue
\mciteSetBstMidEndSepPunct{\mcitedefaultmidpunct}
{\mcitedefaultendpunct}{\mcitedefaultseppunct}\relax
\EndOfBibitem
\bibitem[Yamaguchi \latin{et~al.}(2018)Yamaguchi, Saito, Yoshida, Yamaguchi,
  Yoda, and Seto]{Yamaguchi2018}
Yamaguchi,~T.; Saito,~M.; Yoshida,~K.; Yamaguchi,~T.; Yoda,~Y.; Seto,~M.
  Structural relaxation and viscoelasticity of a higher alcohol with mesoscopic
  structure. \emph{J. Phys. Chem. Lett.} \textbf{2018}, \emph{9},
  298--301\relax
\mciteBstWouldAddEndPuncttrue
\mciteSetBstMidEndSepPunct{\mcitedefaultmidpunct}
{\mcitedefaultendpunct}{\mcitedefaultseppunct}\relax
\EndOfBibitem
\bibitem[Maribo-Mogensen \latin{et~al.}(2013)Maribo-Mogensen, Kontogeorgis, and
  Thomsen]{Maribo2013}
Maribo-Mogensen,~B.; Kontogeorgis,~G.; Thomsen,~K. Modeling of dielectric
  properties of complex fluids with an equation of state. \emph{J. Phys. Chem.
  B} \textbf{2013}, \emph{117}, 3389--3397\relax
\mciteBstWouldAddEndPuncttrue
\mciteSetBstMidEndSepPunct{\mcitedefaultmidpunct}
{\mcitedefaultendpunct}{\mcitedefaultseppunct}\relax
\EndOfBibitem
\bibitem[Angell \latin{et~al.}(2000)Angell, Ngai, McKenna, McMillan, and
  Martin]{Angell2000}
Angell,~C.~A.; Ngai,~K.~L.; McKenna,~G.~B.; McMillan,~P.~F.; Martin,~S.~W.
  Relaxation in glass-forming liquids and amorphous solids. \emph{J. Appl.
  Phys.} \textbf{2000}, \emph{88}, 3113--3157\relax
\mciteBstWouldAddEndPuncttrue
\mciteSetBstMidEndSepPunct{\mcitedefaultmidpunct}
{\mcitedefaultendpunct}{\mcitedefaultseppunct}\relax
\EndOfBibitem
\bibitem[Lomas(2015)]{Lomas2015}
Lomas,~J.~S. 1H NMR spectra of alcohols in hydrogen bonding solvents: DFT/GIAO
  calculations of chemical shifts. \emph{Magnetic Resonance in Chemistry}
  \textbf{2015}, \emph{54}, 28--38\relax
\mciteBstWouldAddEndPuncttrue
\mciteSetBstMidEndSepPunct{\mcitedefaultmidpunct}
{\mcitedefaultendpunct}{\mcitedefaultseppunct}\relax
\EndOfBibitem
\bibitem[Meng \latin{et~al.}(2015)Meng, Guo, Peng, Chen, Wang, Shi, Li, Wang,
  and Jiang]{Meng2015}
Meng,~X.; Guo,~J.; Peng,~J.; Chen,~J.; Wang,~Z.; Shi,~J.; Li,~X.~Z.;
  Wang,~E.-G.; Jiang,~Y. Direct visualization of concerted proton tunnelling in
  a water nanocluster. \emph{Nature Phys.} \textbf{2015}, \emph{11},
  235--239\relax
\mciteBstWouldAddEndPuncttrue
\mciteSetBstMidEndSepPunct{\mcitedefaultmidpunct}
{\mcitedefaultendpunct}{\mcitedefaultseppunct}\relax
\EndOfBibitem
\bibitem[Salna \latin{et~al.}(2016)Salna, Benabbas, Sage, Thor, and
  Champion]{Salna2016}
Salna,~B.; Benabbas,~A.; Sage,~J.~T.; Thor,~J.; Champion,~P.~M.
  Wide-dynamic-range kinetic investigations of deep proton tunnelling in
  proteins. \emph{Nat. Chem.} \textbf{2016}, \emph{8}, 874--880\relax
\mciteBstWouldAddEndPuncttrue
\mciteSetBstMidEndSepPunct{\mcitedefaultmidpunct}
{\mcitedefaultendpunct}{\mcitedefaultseppunct}\relax
\EndOfBibitem
\bibitem[L\"owdin(1963)]{Lowdin1963}
L\"owdin,~P. Proton Tunneling in DNA and its Biological Implications.
  \emph{Rev. Mod. Phys.} \textbf{1963}, \emph{35}, 724--732\relax
\mciteBstWouldAddEndPuncttrue
\mciteSetBstMidEndSepPunct{\mcitedefaultmidpunct}
{\mcitedefaultendpunct}{\mcitedefaultseppunct}\relax
\EndOfBibitem
\bibitem[Wang \latin{et~al.}(2020)Wang, Shan, Chen, Pfeifer, Chen, Ren, and
  Dorn]{Wang2020}
Wang,~E.; Shan,~X.; Chen,~L.; Pfeifer,~T.; Chen,~X.; Ren,~X.; Dorn,~A.
  Ultrafast Proton Transfer Dynamics on the Repulsive Potential of the Ethanol
  Dication: Roaming-Mediated Isomerization versus Coulomb Explosion. \emph{J.
  Phys. Chem. A} \textbf{2020}, \emph{124}, 2785--2791\relax
\mciteBstWouldAddEndPuncttrue
\mciteSetBstMidEndSepPunct{\mcitedefaultmidpunct}
{\mcitedefaultendpunct}{\mcitedefaultseppunct}\relax
\EndOfBibitem
\bibitem[Kremer and Sch\"onhals(2003)Kremer, and Sch\"onhals]{Kremer2003}
Kremer,~F., Sch\"onhals,~A., Eds. \emph{Broadband Dielectric Spectroscopy}, 1st
  ed.; Springer: Berlin, 2003\relax
\mciteBstWouldAddEndPuncttrue
\mciteSetBstMidEndSepPunct{\mcitedefaultmidpunct}
{\mcitedefaultendpunct}{\mcitedefaultseppunct}\relax
\EndOfBibitem
\bibitem[Yomogida \latin{et~al.}(2010)Yomogida, Sato, Nozaki, Mishina, and
  Nakahara]{Yomogida2010}
Yomogida,~Y.; Sato,~Y.; Nozaki,~R.; Mishina,~T.; Nakahara,~J. Dielectric study
  of normal alcohols with THz time-domain spectroscopy. \emph{J. Mol. Liq.}
  \textbf{2010}, \emph{154}, 31--35\relax
\mciteBstWouldAddEndPuncttrue
\mciteSetBstMidEndSepPunct{\mcitedefaultmidpunct}
{\mcitedefaultendpunct}{\mcitedefaultseppunct}\relax
\EndOfBibitem
\bibitem[Sarkar \latin{et~al.}(2017)Sarkar, Saha, Banerjee, Mukherjee, and
  Mandal]{Sarkar2017}
Sarkar,~S.; Saha,~D.; Banerjee,~S.; Mukherjee,~A.; Mandal,~P. Broadband
  terahertz dielectric spectroscopy of alcohols. \emph{Chem. Phys. Lett.}
  \textbf{2017}, \emph{678}, 65--71\relax
\mciteBstWouldAddEndPuncttrue
\mciteSetBstMidEndSepPunct{\mcitedefaultmidpunct}
{\mcitedefaultendpunct}{\mcitedefaultseppunct}\relax
\EndOfBibitem
\bibitem[Prego \latin{et~al.}(2000)Prego, Cabeza, Carballo, Franjo, and
  Jimenez]{Prego2000}
Prego,~M.; Cabeza,~O.; Carballo,~E.; Franjo,~C.~F.; Jimenez,~E. Measurement and
  interpretation of the electrical conductivity of 1-alcohols from 273 K to 333
  K. \emph{J. Mol. Liq.} \textbf{2000}, \emph{89}, 233--238\relax
\mciteBstWouldAddEndPuncttrue
\mciteSetBstMidEndSepPunct{\mcitedefaultmidpunct}
{\mcitedefaultendpunct}{\mcitedefaultseppunct}\relax
\EndOfBibitem
\bibitem[Prego \latin{et~al.}(2003)Prego, Rilo, Carballo, Franjo, Jim\'enez,
  and Cabeza]{Prego2003}
Prego,~M.; Rilo,~E.; Carballo,~E.; Franjo,~C.; Jim\'enez,~E.; Cabeza,~O.
  Electrical conductivity data of alkanols from 273 to 333 K. \emph{J. Mol.
  Liq.} \textbf{2003}, \emph{102}, 83--91\relax
\mciteBstWouldAddEndPuncttrue
\mciteSetBstMidEndSepPunct{\mcitedefaultmidpunct}
{\mcitedefaultendpunct}{\mcitedefaultseppunct}\relax
\EndOfBibitem
\bibitem[Pratt and Wakeham(1977)Pratt, and Wakeham]{Pratt1977}
Pratt,~K.~C.; Wakeham,~W. Self-Diffusion in Water and Monohydric Alcohols.
  \emph{J. Chem. Soc., Faraday Trans.} \textbf{1977}, \emph{73},
  997--1002\relax
\mciteBstWouldAddEndPuncttrue
\mciteSetBstMidEndSepPunct{\mcitedefaultmidpunct}
{\mcitedefaultendpunct}{\mcitedefaultseppunct}\relax
\EndOfBibitem
\bibitem[Robb(1982)]{robb1982}
Robb,~I.~D., Ed. \emph{Microemulsions}, 1st ed.; Plenum: New York, 1982\relax
\mciteBstWouldAddEndPuncttrue
\mciteSetBstMidEndSepPunct{\mcitedefaultmidpunct}
{\mcitedefaultendpunct}{\mcitedefaultseppunct}\relax
\EndOfBibitem
\bibitem[Ugur \latin{et~al.}(2014)Ugur, Marion, Parant, Jensen, and
  Monard]{Ugur2014}
Ugur,~I.; Marion,~A.; Parant,~S.; Jensen,~J.~H.; Monard,~G. Rationalization of
  the pKa values of alcohols and thiols using atomic charge descriptors and its
  application to the prediction of amino acid pKa's. \emph{J. Chem. Inform. and
  Modeling} \textbf{2014}, \emph{54}, 2200--13\relax
\mciteBstWouldAddEndPuncttrue
\mciteSetBstMidEndSepPunct{\mcitedefaultmidpunct}
{\mcitedefaultendpunct}{\mcitedefaultseppunct}\relax
\EndOfBibitem
\bibitem[Miller and Manolopoulos(2005)Miller, and Manolopoulos]{Miller2005}
Miller,~T.~F.; Manolopoulos,~D.~E. Quantum diffusion in liquid water from ring
  polymer molecular dynamics. \emph{J. Chem. Phys.} \textbf{2005}, \emph{123},
  154504--10\relax
\mciteBstWouldAddEndPuncttrue
\mciteSetBstMidEndSepPunct{\mcitedefaultmidpunct}
{\mcitedefaultendpunct}{\mcitedefaultseppunct}\relax
\EndOfBibitem
\bibitem[Habershon \latin{et~al.}(2009)Habershon, Markland, and
  Manolopoulos]{Habershon2009}
Habershon,~S.; Markland,~T.~E.; Manolopoulos,~D.~E. Competing quantum effects
  in the dynamics of a flexible water model. \emph{J. Chem. Phys.}
  \textbf{2009}, \emph{131}, 024501--11\relax
\mciteBstWouldAddEndPuncttrue
\mciteSetBstMidEndSepPunct{\mcitedefaultmidpunct}
{\mcitedefaultendpunct}{\mcitedefaultseppunct}\relax
\EndOfBibitem
\bibitem[Novikov and Sokolov(2013)Novikov, and Sokolov]{Novikov2013}
Novikov,~V.~N.; Sokolov,~A.~P. Role of quantum effects in the glass transition.
  \emph{Phys. Rev. Lett.} \textbf{2013}, \emph{110}, 065701--5\relax
\mciteBstWouldAddEndPuncttrue
\mciteSetBstMidEndSepPunct{\mcitedefaultmidpunct}
{\mcitedefaultendpunct}{\mcitedefaultseppunct}\relax
\EndOfBibitem
\bibitem[Bakker and Nienhuys(2002)Bakker, and Nienhuys]{Bakker2002}
Bakker,~H.~J.; Nienhuys,~H.-K. Delocalization of Protons in Liquid Water.
  \emph{Science} \textbf{2002}, \emph{297}, 587--590\relax
\mciteBstWouldAddEndPuncttrue
\mciteSetBstMidEndSepPunct{\mcitedefaultmidpunct}
{\mcitedefaultendpunct}{\mcitedefaultseppunct}\relax
\EndOfBibitem
\bibitem[Artemov \latin{et~al.}(2020)Artemov, Uykur, Roh, Pronin, Ouerdane, and
  Dressel]{Artemov2020}
Artemov,~V.~G.; Uykur,~E.; Roh,~S.; Pronin,~A.; Ouerdane,~H.; Dressel,~M.
  Revealing excess protons in the infrared spectrum of liquid water.
  \emph{Scientific Reports} \textbf{2020}, \emph{10}, 11320--7\relax
\mciteBstWouldAddEndPuncttrue
\mciteSetBstMidEndSepPunct{\mcitedefaultmidpunct}
{\mcitedefaultendpunct}{\mcitedefaultseppunct}\relax
\EndOfBibitem
\bibitem[Mazur \latin{et~al.}(2015)Mazur, Bonn, and Hunger]{Mazur2015}
Mazur,~K.; Bonn,~M.; Hunger,~J. Hydrogen bond dynamics in primary alcohols: A
  femtosecond infrared study. \emph{J. Phys. Chem. B} \textbf{2015},
  \emph{119}, 1558--1566\relax
\mciteBstWouldAddEndPuncttrue
\mciteSetBstMidEndSepPunct{\mcitedefaultmidpunct}
{\mcitedefaultendpunct}{\mcitedefaultseppunct}\relax
\EndOfBibitem
\bibitem[Markovitch \latin{et~al.}(2008)Markovitch, Chen, Izvekov, Paesani,
  Voth, and Agmon]{Markovitch2008}
Markovitch,~O.; Chen,~H.; Izvekov,~S.; Paesani,~F.; Voth,~G.~A.; Agmon,~N.
  Special pair dance and partner selection: Elementary steps in proton
  transport in liquid water. \emph{J. Phys. Chem. B} \textbf{2008}, \emph{112},
  9456--9466\relax
\mciteBstWouldAddEndPuncttrue
\mciteSetBstMidEndSepPunct{\mcitedefaultmidpunct}
{\mcitedefaultendpunct}{\mcitedefaultseppunct}\relax
\EndOfBibitem
\bibitem[Pietropaolo \latin{et~al.}(2008)Pietropaolo, Senesi, Andreani, Botti,
  Ricci, and Bruni]{Pietropaolo2008}
Pietropaolo,~A.; Senesi,~R.; Andreani,~C.; Botti,~A.; Ricci,~M.~A.; Bruni,~F.
  Excess of proton mean kinetic energy in supercooled water. \emph{Phys. Rev.
  Lett.} \textbf{2008}, \emph{100}, 127802--4\relax
\mciteBstWouldAddEndPuncttrue
\mciteSetBstMidEndSepPunct{\mcitedefaultmidpunct}
{\mcitedefaultendpunct}{\mcitedefaultseppunct}\relax
\EndOfBibitem
\bibitem[Lin \latin{et~al.}(2010)Lin, Morrone, Car, and Parrinello]{Lin2010}
Lin,~L.; Morrone,~J.; Car,~R.; Parrinello,~M. Displaced path integral
  formulation for the momentum distribution of quantum particles. \emph{Phys.
  Rev. Lett.} \textbf{2010}, \emph{105}, 110602--4\relax
\mciteBstWouldAddEndPuncttrue
\mciteSetBstMidEndSepPunct{\mcitedefaultmidpunct}
{\mcitedefaultendpunct}{\mcitedefaultseppunct}\relax
\EndOfBibitem
\bibitem[Geissler \latin{et~al.}(2001)Geissler, Dellago, Chandler, Hutter, and
  Parrinello]{Geissler2001}
Geissler,~P.~L.; Dellago,~C.; Chandler,~D.; Hutter,~J.; Parrinello,~M.
  Autoionization in Liquid Water. \emph{Science} \textbf{2001}, \emph{291},
  2121--2124\relax
\mciteBstWouldAddEndPuncttrue
\mciteSetBstMidEndSepPunct{\mcitedefaultmidpunct}
{\mcitedefaultendpunct}{\mcitedefaultseppunct}\relax
\EndOfBibitem
\bibitem[Hassanali \latin{et~al.}(2013)Hassanali, Giberti, Cuny, K\"uhne, and
  Parrinello]{Hassanali2013}
Hassanali,~A.; Giberti,~F.; Cuny,~J.; K\"uhne,~T.; Parrinello,~M. Proton
  transfer through the water gossamer. \emph{PNAS} \textbf{2013}, \emph{110},
  13723--13728\relax
\mciteBstWouldAddEndPuncttrue
\mciteSetBstMidEndSepPunct{\mcitedefaultmidpunct}
{\mcitedefaultendpunct}{\mcitedefaultseppunct}\relax
\EndOfBibitem
\bibitem[Hawlicka(1995)]{Hawlicka1995}
Hawlicka,~E. Self-diffusion in multicomponent liquid systems. \emph{Chem. Soc.
  Rev.} \textbf{1995}, \emph{24}, 367--377\relax
\mciteBstWouldAddEndPuncttrue
\mciteSetBstMidEndSepPunct{\mcitedefaultmidpunct}
{\mcitedefaultendpunct}{\mcitedefaultseppunct}\relax
\EndOfBibitem
\bibitem[Balberg \latin{et~al.}(1984)Balberg, Binenbaum, and
  Wagner]{Balberg1984}
Balberg,~I.; Binenbaum,~N.; Wagner,~N. Percolation thresholds in the
  three-dimensional sticks system. \emph{Phys. Rev. Lett.} \textbf{1984},
  \emph{52}, 1465--1468\relax
\mciteBstWouldAddEndPuncttrue
\mciteSetBstMidEndSepPunct{\mcitedefaultmidpunct}
{\mcitedefaultendpunct}{\mcitedefaultseppunct}\relax
\EndOfBibitem
\bibitem[Davidson(1957)]{Davidson1957}
Davidson,~D.~W. The dielectric properties of methanol and methanol-d.
  \emph{Can. J. Chem.} \textbf{1957}, \emph{35}, 458--473\relax
\mciteBstWouldAddEndPuncttrue
\mciteSetBstMidEndSepPunct{\mcitedefaultmidpunct}
{\mcitedefaultendpunct}{\mcitedefaultseppunct}\relax
\EndOfBibitem
\bibitem[Sedov and Solomonov(2012)Sedov, and Solomonov]{Sedov2012}
Sedov,~I.; Solomonov,~B. Gibbs free energy of hydrogen bonding of aliphatic
  alcohols with liquid water at 298 K. \emph{Fluid Phase Equilibria}
  \textbf{2012}, \emph{315}, 16--20\relax
\mciteBstWouldAddEndPuncttrue
\mciteSetBstMidEndSepPunct{\mcitedefaultmidpunct}
{\mcitedefaultendpunct}{\mcitedefaultseppunct}\relax
\EndOfBibitem
\bibitem[Th\"amer \latin{et~al.}(2015)Th\"amer, Marco, Ramasesha, Mandal, and
  Tokmakoff]{Thamer2015}
Th\"amer,~M.; Marco,~L.~D.; Ramasesha,~K.; Mandal,~A.; Tokmakoff,~A. Ultrafast
  2D IR spectroscopy of the excess proton in liquid water. \emph{Science}
  \textbf{2015}, \emph{350}, 78--82\relax
\mciteBstWouldAddEndPuncttrue
\mciteSetBstMidEndSepPunct{\mcitedefaultmidpunct}
{\mcitedefaultendpunct}{\mcitedefaultseppunct}\relax
\EndOfBibitem
\bibitem[Gainaru \latin{et~al.}(2011)Gainaru, Kastner, Mayr, Lunkenheimer,
  Schildmann, Weber, Hiller, Loidl, and B\"ohmer]{Gainaru2011}
Gainaru,~C.; Kastner,~S.; Mayr,~F.; Lunkenheimer,~P.; Schildmann,~S.;
  Weber,~H.; Hiller,~W.; Loidl,~A.; B\"ohmer,~R. Hydrogen-bond equilibria and
  lifetimes in a monohydroxy alcohol. \emph{Phys. Rev. Lett.} \textbf{2011},
  \emph{107}, 118304--5\relax
\mciteBstWouldAddEndPuncttrue
\mciteSetBstMidEndSepPunct{\mcitedefaultmidpunct}
{\mcitedefaultendpunct}{\mcitedefaultseppunct}\relax
\EndOfBibitem
\bibitem[Grechko \latin{et~al.}(2018)Grechko, Hasegawa, D’Angelo, Ito,
  Turchinovich, Nagata, and Bonn]{Grechko2018}
Grechko,~M.; Hasegawa,~T.; D’Angelo,~F.; Ito,~H.; Turchinovich,~D.;
  Nagata,~Y.; Bonn,~M. Coupling between intra- and intermolecular motions in
  liquid water revealed by two-dimensional terahertz-infrared-visible
  spectroscopy. \emph{Nat. Comm.} \textbf{2018}, \emph{9}, 885--8\relax
\mciteBstWouldAddEndPuncttrue
\mciteSetBstMidEndSepPunct{\mcitedefaultmidpunct}
{\mcitedefaultendpunct}{\mcitedefaultseppunct}\relax
\EndOfBibitem
\bibitem[Guo \latin{et~al.}(2003)Guo, Luo, Augustsson, Kashtanov, Rubensson,
  Shuh, Ågren, and Nordgren]{Guo2003}
Guo,~J.-H.; Luo,~Y.; Augustsson,~A.; Kashtanov,~S.; Rubensson,~J.-E.;
  Shuh,~D.~K.; Ågren,~H.; Nordgren,~J. Molecular structure of alcohol-water
  mixtures. \emph{Phys. Rev. Lett.} \textbf{2003}, \emph{91}, 157401--4\relax
\mciteBstWouldAddEndPuncttrue
\mciteSetBstMidEndSepPunct{\mcitedefaultmidpunct}
{\mcitedefaultendpunct}{\mcitedefaultseppunct}\relax
\EndOfBibitem
\bibitem[Gabriel \latin{et~al.}(2018)Gabriel, Pabst, Helbling, B\"ohmer, and
  Blochowicz]{Gabriel2018}
Gabriel,~J.; Pabst,~F.; Helbling,~A.; B\"ohmer,~T.; Blochowicz,~T. Nature of
  the Debye process in monohydroxy alcohols: 5-methyl-2-hexanol investigated by
  depolarized light scattering and dielectric spectroscopy. \emph{Phys. Rev.
  Lett.} \textbf{2018}, \emph{121}, 035501--5\relax
\mciteBstWouldAddEndPuncttrue
\mciteSetBstMidEndSepPunct{\mcitedefaultmidpunct}
{\mcitedefaultendpunct}{\mcitedefaultseppunct}\relax
\EndOfBibitem
\bibitem[Singh and Richert(2012)Singh, and Richert]{Singh2012}
Singh,~L.; Richert,~R. Watching hydrogen-bonded structures in an alcohol
  convert from rings to chains. \emph{Phys. Rev. Lett.} \textbf{2012},
  \emph{109}, 167802--5\relax
\mciteBstWouldAddEndPuncttrue
\mciteSetBstMidEndSepPunct{\mcitedefaultmidpunct}
{\mcitedefaultendpunct}{\mcitedefaultseppunct}\relax
\EndOfBibitem
\bibitem[Frenkel(1946)]{Frenkel1946}
Frenkel,~J. \emph{Kinetic theory of liquids}, 1st ed.; Clarendon Press: Oxford,
  1946\relax
\mciteBstWouldAddEndPuncttrue
\mciteSetBstMidEndSepPunct{\mcitedefaultmidpunct}
{\mcitedefaultendpunct}{\mcitedefaultseppunct}\relax
\EndOfBibitem
\end{mcitethebibliography}

\end{document}